\documentclass[conference]{IEEEtran}
\IEEEoverridecommandlockouts
\usepackage{cite}
\usepackage{amsmath,amssymb,amsfonts}
\usepackage{algorithmic}
\usepackage{graphicx}
\usepackage{textcomp}
\usepackage{xcolor}
\usepackage{upgreek}
\usepackage[table]{xcolor}
\usepackage{relsize}
\usepackage[caption=false,font=footnotesize]{subfig}

\usepackage{colortbl}
\usepackage{makecell}

\def\BibTeX{{\rm B\kern-.05em{\sc i\kern-.025em b}\kern-.08em
    T\kern-.1667em\lower.7ex\hbox{E}\kern-.125emX}}
\begin{document}

\title{A 32-Channel 3.53-$\mu$W per Channel Brain-machine Interface SoC Featuring Dual-threshold Delta-modulation, In-memory Spike Detection and Bi-SNN based Motor Decoding\\

\thanks{Acknowledgements This work was supported by RGC of the HKSAR, China (Project No. CityU 11200922).}
}

\author{
    \IEEEauthorblockN{Ye Ke\textsuperscript{†,1}, Zhengnan Fu\textsuperscript{†,1}, Pao-Sheng Vincent Sun\textsuperscript{1}, An Guo\textsuperscript{1,2}, Shuai Dong\textsuperscript{1}, Junyi Yang\textsuperscript{1},}
    \IEEEauthorblockN{Yahan Yang\textsuperscript{1}, Abdelrahman B. M. Eldaly\textsuperscript{1}, Xin Si\textsuperscript{2}, Leanne Chan\textsuperscript{1}, Arindam Basu\textsuperscript{1}}
    \IEEEauthorblockA{\textit{\textsuperscript{†} Equally Credited Authors (ECAs)}\\
    \textit{\textsuperscript{1} City University of Hong Kong, Hong Kong SAR, China}\\
    \textit{\textsuperscript{2 }Southeast University, Nanjing, China}}
}
\maketitle

\begin{abstract}
With the scaling of sensor channel counts, iBMI systems face challenges in frontend data sensing and on-implant data processing.
This work presents a 32-channel fully event-based iBMI SoC in 65nm CMOS for an efficient neuromorphic signal processing pipeline. 
The SoC integrates a 32-channel dual-threshold delta modulation (DTDM) frontend array that provides up to 26× data compression at the frontend, an in-memory computing (IMC) spike detector (SPD) for efficient in-pixel spike detection, and a bipolar LIF-based spiking neural network (Bi-SNN) decoder for on-chip motor intention decoding (MID).
Consuming only 3.53 $\mu W$ per channel and achieving ~0.62 decoding $R^2$ with a compact 0.034 $mm^2$ per-channel area, the chip enables high-efficiency signal recording, processing, and decoding for implantable devices.
\end{abstract}

\section{Introduction}
\begin{figure}[t]

    \centering
    \includegraphics[width=1\linewidth]{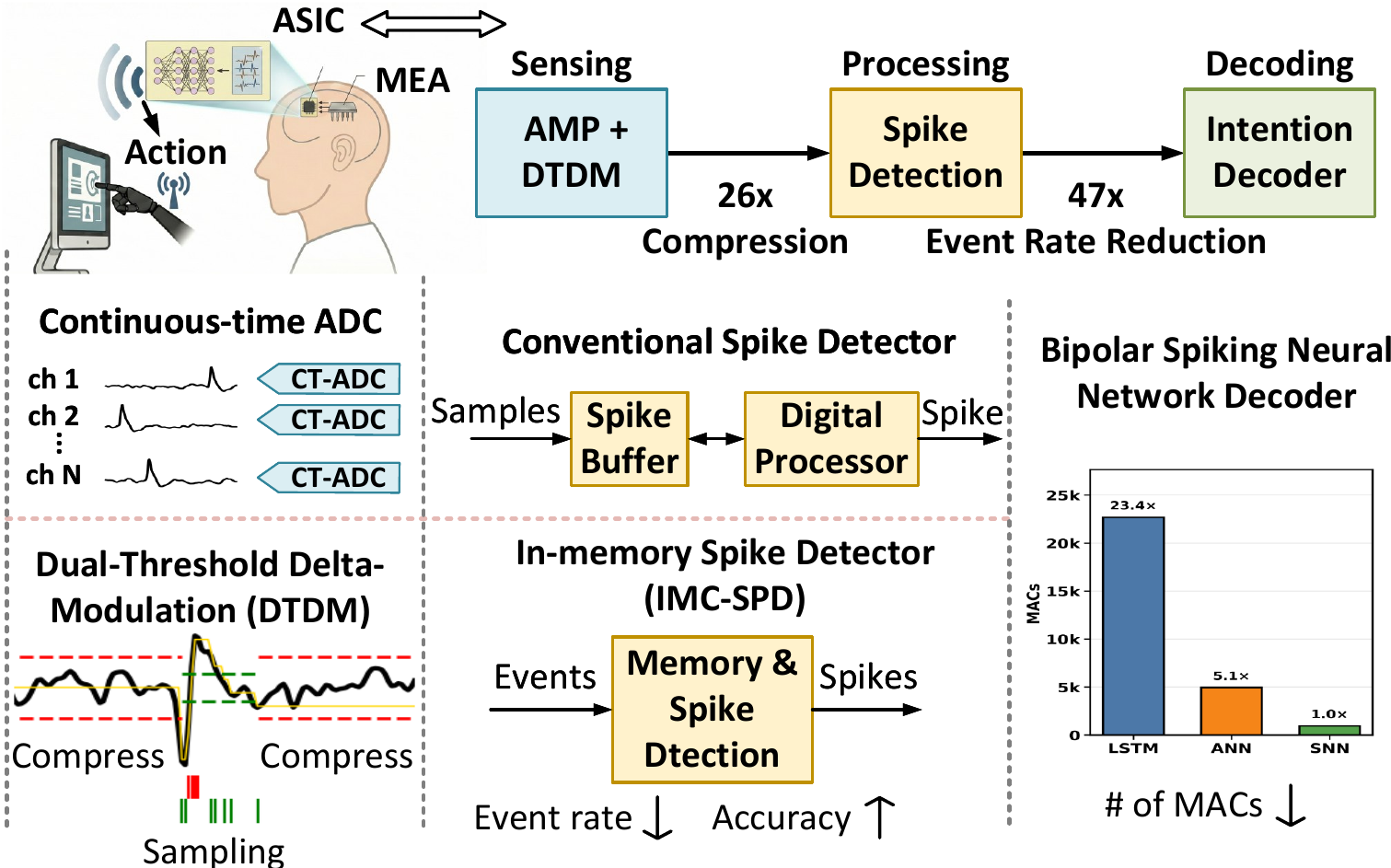}
    \caption{Design challenges in high-density wireless iBMIs system and proposed approaches.}
    \label{fig_intro}
    \vspace{-10pt}
\end{figure}

As intracortical brain–machine interfaces (iBMIs) evolve toward higher channel counts, the increasing data rate from the frontend exceeds the wireless transmission capacity of an implantable device.
As a step towards next-generation wireless and high-density iBMI systems, an end-to-end on-implant iBMI pipeline with signal acquisition, processing, and decoding, as shown in Fig.\ref {fig_intro}, is a promising approach to reduce data rate from neural implants. 
The on-implant pipeline aims to compress and process neural data at each step and to transmit only minimal control information off-chip.
However, conventional iBMI systems face bottlenecks throughout the workflow. At the frontend, conventional Nyquist ADCs perform dense, uniform sampling of the sparse neural signal, resulting in high energy consumption and increased data rates. 
Consequently, detecting action potentials from raw neural recordings requires a large sample buffer and heavy computation, and decoding intentions from neural activity also demands dense multiply-and-accumulate (MAC) operations. 

To address these emerging challenges for neural implants, we present a 32-channel end-to-end iBMI system-on-chip (SoC) with a fully event-based neuromorphic workflow (Fig \ref{fig_structure}). 
At the frontend, we implemented an analog-to-spike converter (ASC) with a dual-threshold delta-modulation (DTDM) frontend. 
While existing event-based sampling techniques \cite{chenNeuronInspired00032mm2138mW2024} have enabled considerable data compression by exploiting neural activity sparsity, the threshold setting for event-based sampling faces a trade-off between compression and robustness to noise. 
Our proposed dual-threshold sampling mechanism uses a coarse threshold for improved noise rejection and data compression, along with a fine threshold for accurate signal reconstruction.
Although DTDM had improved noise immunity, there are still spurious events generated by noise which can reduce accuracy of downstream decoders. 
Hence, we designed an in-pixel in-memory computing neural spike detector (IMC-SPD) to reduce the number of false events due to noise and concomitantly, the AER handshakes from the frontend array by 47-101×.
The IMC-SPD bitcell leverages the event-based nature of the frontend by storing temporal events directly in eDRAM and SRAM cells, and then detecting neural spikes via in-memory logic, reducing the buffer size by 9.5×.
Our iBMI SoC also features a lightweight AER-compatible bipolar spiking neural network (Bi-SNN) decoder optimized for motor intention decoding (MID). The event-based nature of SNN architecture reduced the number of MACs by a factor of 23.4× while maintaining comparable MID performance compared with ANN neural decoders.

\section{Proposed SOC Architecture}
\begin{figure}[t]
    \centering
    \vspace{-10pt}
    \includegraphics[width=\linewidth]{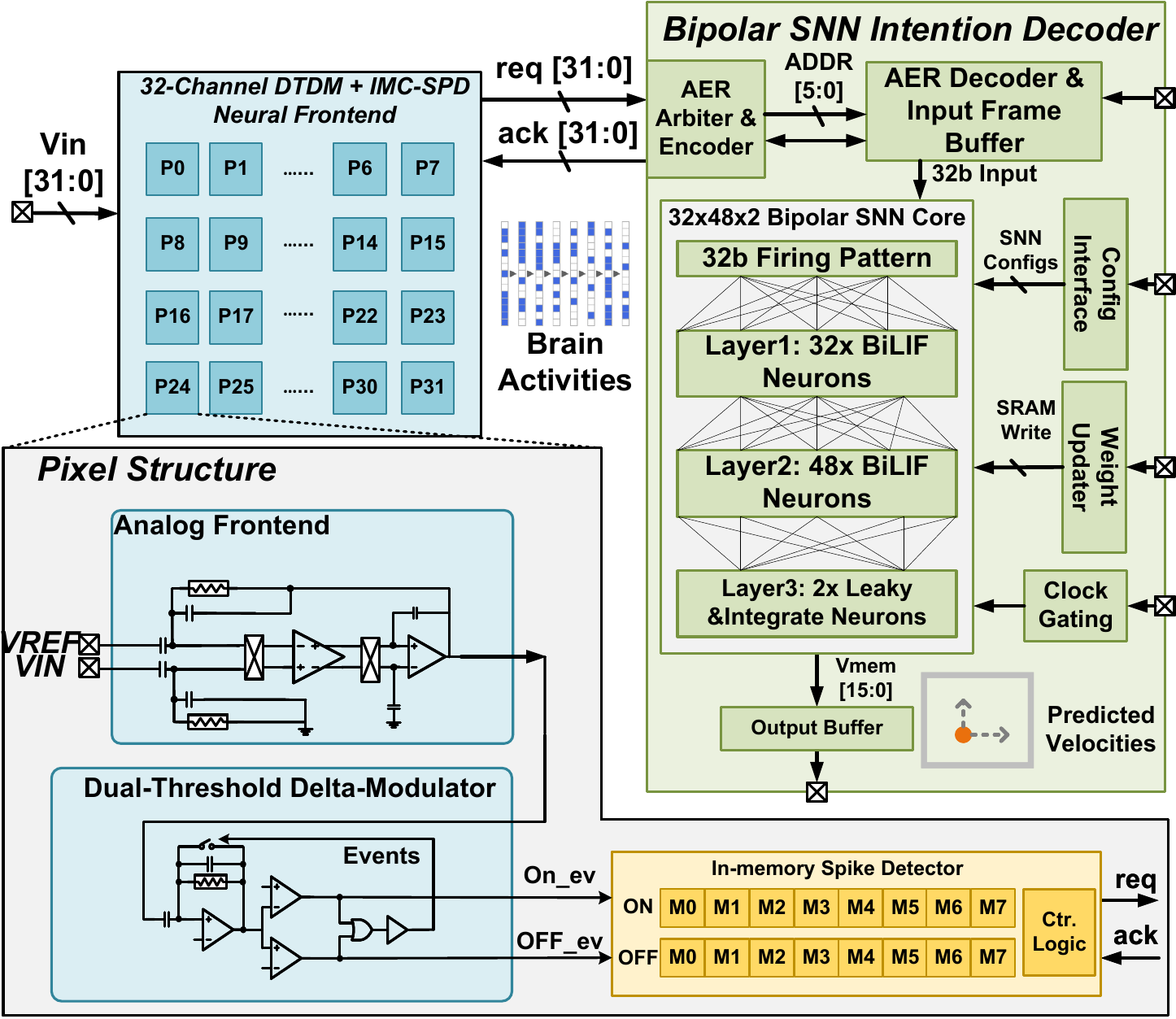}
    \caption{Top-level SoC architecture, neural sensing, processing pixel structure and neural decoder block diagram.}
    \label{fig_structure}
    \vspace{-10pt}
\end{figure}

Figure \ref{fig_structure} presents the proposed fully event-based iBMI SoC architecture, including a 32-channel neural sensing and processing array, an AER interface, and an on-chip Bi-SNN MID decoder. 
At the frontend, each neural sensing pixel in the 4x8 array comprises a low-noise neural amplifier and a DTDM modulator.
The on/off events from the modulator are processed by the IMC-SPD cells and control logic within the pixel to detect the action potentials from events.
In the presence of spikes, the pixel sends requests to the AER interface, and the spikes are stored in the SNN input buffer for MID. 
The decoder incorporates a 32-48-2 BI-SNN core that takes brain-activity pattern frames as input and outputs predicted motor velocities.
\subsection{Integrated Architecture of Analog to Spike Converter}
\subsubsection{Clamped-T Capacitor Amplifier With Chopper}

The AFE employs a chopping-based topology to reduce the effect of $1/f$ noise by modulating the input signal into a higher-frequency band and demodulating the amplified output back to baseband, thereby avoiding the need for large transistors. The clamped-T topology replaces the conventional feedback capacitor with a T-capacitor network and clamping diodes, reducing its effective value to $Cu/(2N+1)$. Combined with input capacitors of magnitude  $M×Cu$, this achieves a high gain of $M×(2N+1)$ while reducing total capacitance by 64\% for M= 30 and N = 1. The AFE further incorporates a two-stage current-reuse OTA that achieves 2$\times$ the open-loop gain at 0.5$\times$ the power consumption of conventional single-stage designs. 

\begin{figure}[t]
    \centering
    \vspace{-10pt}
    \includegraphics[width=1\linewidth]{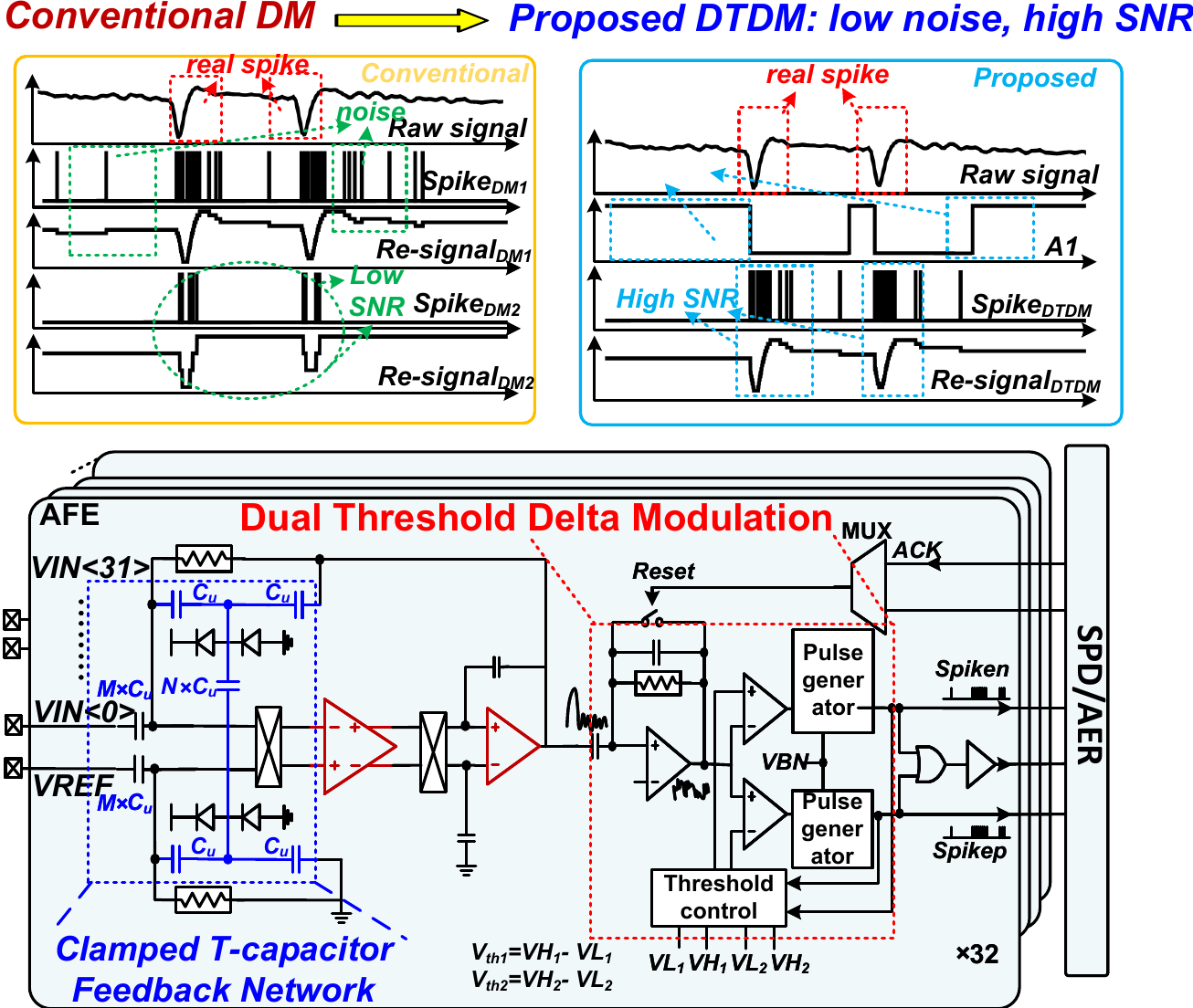}
    \caption{Architecture and Circuit Implementation of the 32-Channel Chopping-Based AFE with Clamped-T-Capacitor Feedback and Dual delta modulation.}
    \label{fig_AFE}
    \vspace{-10pt}
\end{figure}

\subsubsection{Dual-Threshold Delta-Modulation Quantizer}

Following the AFE, the DTDM circuit serves as an integrated stage for quantization and event-based data compression. Unlike conventional Delta Modulation (DM) with a single fixed threshold, the DTDM employs an adaptive dual-threshold mechanism: the AFE output is continuously compared against upper boundary $ V_{H}$ and lower boundary$ V_{L}$. Once the change in signal crosses a predefined coarse threshold $V_{th1}=V_{H1}-V_{L1}$, a binary event (`spiken' or `spikep') is generated, and the amplifier is reset. Subsequently, the dual-threshold controller uses a fine threshold window $V_{th2}=V_{H2}-V_{L2}$ to improve the tracking rate; conversely, when no event is detected, the quantizer uses the coarse threshold window $ V_{th1}$ to suppress noise induced events.

The DTDM is implemented as a switched-capacitor circuit, comprising a self-biased low-power comparator and a digital state machine. To balance response speed and noise stability, the temporal window is set to 1.5 ms.

\subsection{In-memory Spike Detector (SPD)}
\begin{figure}[t]
    \centering
    \includegraphics[width=\linewidth]{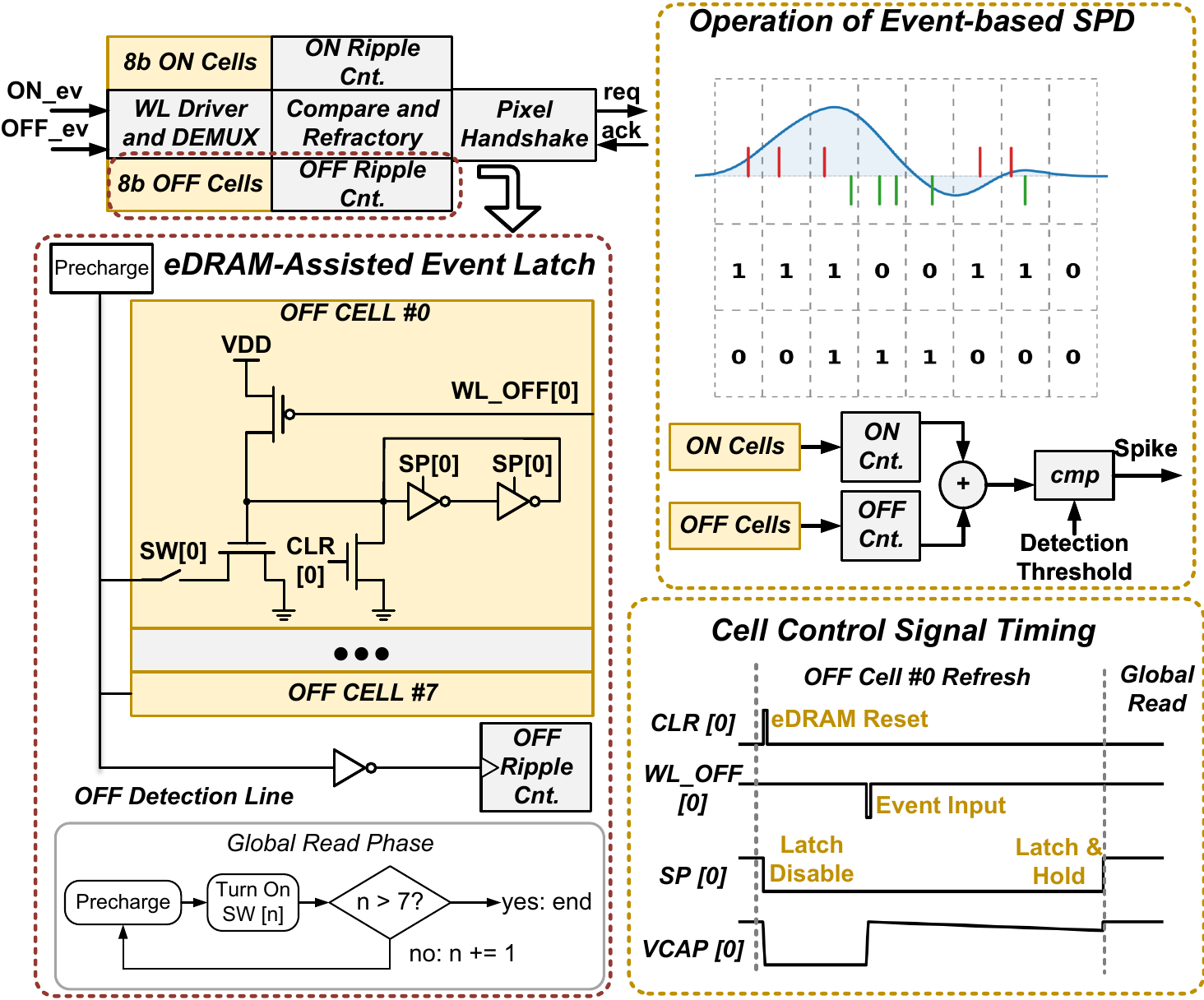}
    \caption{Circuit implementation of eDRAM-assisted event latch for in-memory spike detection.}
    \label{fig_HRAM}
    \vspace{-10pt}
\end{figure}

 Figure \ref{fig_HRAM} depicts the details of the proposed IMC-SPD. 
 The proposed detector implements event-based spike detection through an eDRAM-assisted event latch. There are 8 unit cells for the processing of ON and OFF events per channel. Each cell processes events for $T_s$ = 125 $\mu$s, resulting in a total sliding window size of 1 ms. The current cell is selected by a pointer in a round robin fashion. During a unit cell update, the latch would be disabled through current starving, and the incoming event would charge the eDRAM capacitor for sample time $T_s$. The capacitor voltage would be binarized and stored by the latch at the end of each $T_s$ to prevent leakage induced information loss in the eDRAM. 
 
 After each bitcell update, the sum of all bitcells is read out through a shared detection line. The detection line is precharged and then connected to the eDRAM storage transistor, and will be discharged if the current latch stores a $1$. In this way, the population count of the bitcells is converted and mapped into the number of pulses occurring on the detection line and can be stored in a ripple counter. 
 The ripple counter readout is compared with a configurable digital threshold to give the detection results.
 Compared with prior off-array iBMI AER spike detectors\cite{ke1024Channel08V239nW2025}, the in-pixel integration of IMC-SPD managed to not only filter noise events but also reduce the AER communication overhead, resulting in better scalability. 

\subsection{Bipolar Spiking Neural Network Decoder}
\begin{figure}[t]
    \centering
    \includegraphics[width=\linewidth]{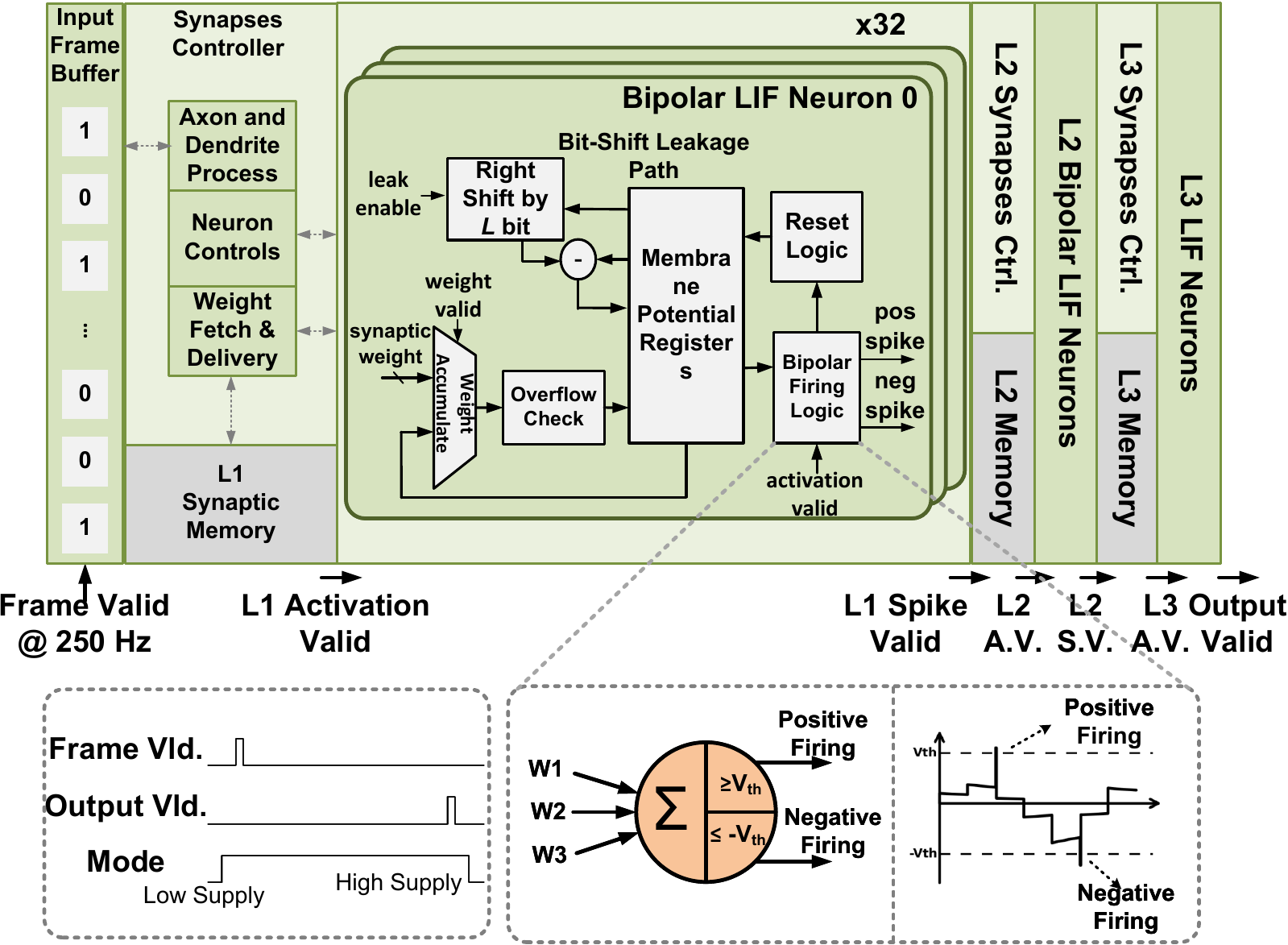}
    \caption{Bipolar SNN intention decoder and hardware implementation of the bipolar LIF neuron.}
    \label{fig_SNN}
    \vspace{-10pt}
\end{figure}

We implemented a dedicated light-weight Bi-SNN core for end-to-end motor intention decoding from the neural firing patterns as depicted in Fig. \ref{fig_SNN}.
The biological spikes detected events identified by IMC-SPD are stored in a compact 32-b input frame buffer that represents a per-channel firing pattern sampled at 250 Hz. 
When the frame valid signal arrives, the compute signal propagates through the network. A synaptic controller fetches 4-b signed weights from the synaptic memory and passes them to the bipolar LIF neurons. 
The bipolar LIF neurons are optimized for power efficiency. We implemented a bit-shift leakage path and a subtractor to mimic an exponential $ V _ {mem} $ leakage behavior with minimal resources.
On the output side, the neuron uses bipolar firing to better encode the temporal dynamics of the membrane potential by ternary activations.
Due to the sparse nature of neural activity, the SNN core can be configured to operate in either idle or inference mode. In idle mode, the core uses a lower power supply and only acknowledges and buffers AER input frames to reduce switching.


\begin{figure}[b]
    \centering
    \vspace{-10pt}
    \includegraphics[width=1\linewidth]{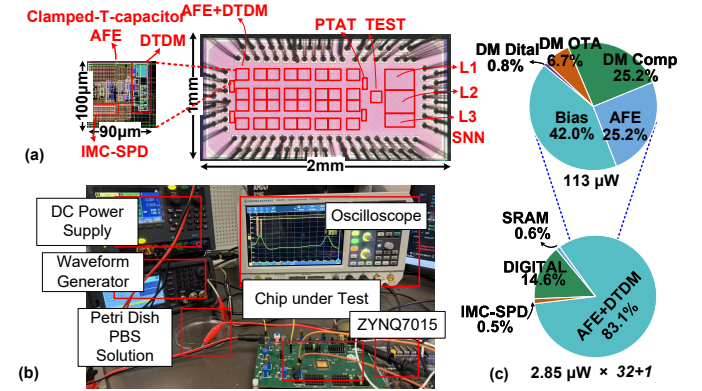}
    \caption{(a) Chip Microscopic diagram, (b) Experimental setup, (c) Power breakdown.}
    \label{fig_combined_result}
\end{figure}

\section{Measurement Results}
\label{sec:experimental_results}
\subsection{Measurement Setup}

The proposed iBMI SoC was fabricated in a 65-nm CMOS process and packaged for system-level measurements. Broadband intracortical recordings from a public dataset \cite{o_doherty_2020_3854034} were replayed via an AWG and attenuated to $200\,\mu\text{V}_{\text{PP}}$ before being injected into a saline solution to emulate the electrode-tissue interface impedance. This setup allowed for a comprehensive assessment of the entire signal chain, from analog acquisition and feature extraction to the on-chip Bi-SNN (32-48-2) decoding of 2D kinematic velocities. All measurements were recorded at a 1.2-V supply and 25$^{\circ}$C.

\subsection{Analog To Spike Converter Performance}
\begin{figure}[t]
    \centering
    \includegraphics[width=1\linewidth]{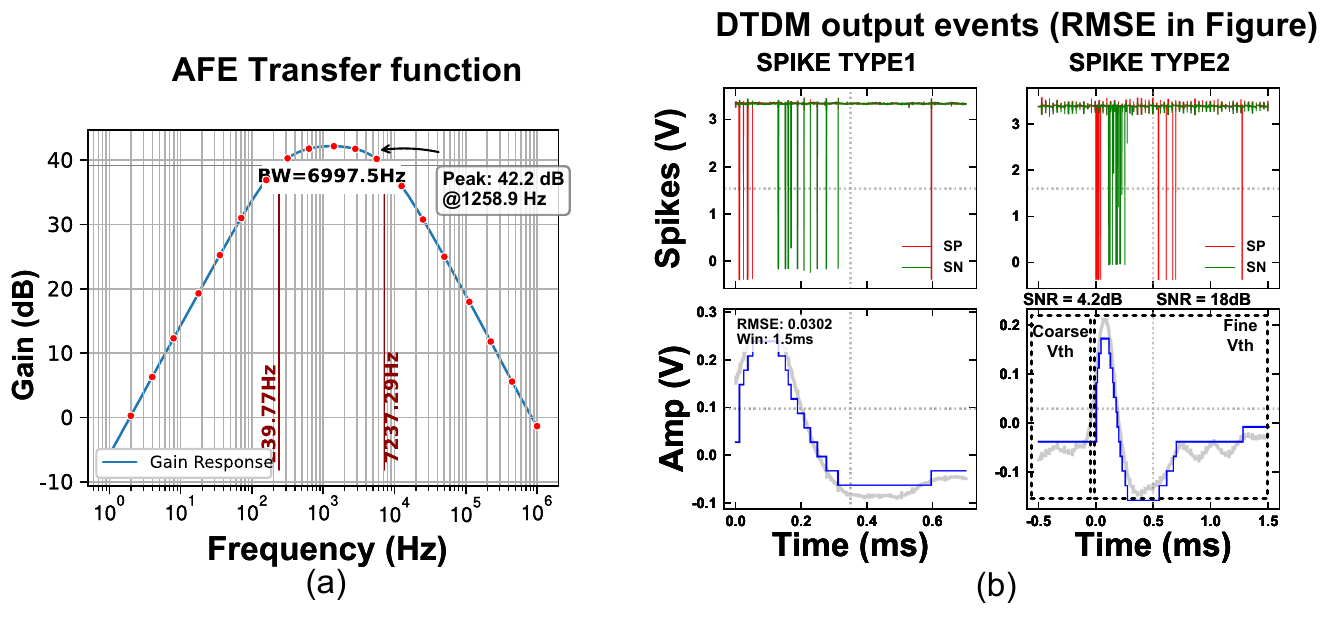}
    \caption{(a) AFE transfer function, (b) DTDM output and wave reconstruction}
    \label{fig_Analogresult}
    \vspace{-10pt}
\end{figure}

The transfer function of the proposed AFE exhibits a mid-band gain of 42 dB with a passband ripple of less than 1 dB, as shown in Fig. \ref{fig_Analogresult}(a). The output spectrum in demonstrates the AFE performance with a measured SNDR of 29.5 dB and an SFDR of 34.0 dB. The measured input-referred noise spectrum in achieves 7.56 $\mu$V$_{rms}$ integrated noise after chopping over the 239 Hz--7.2 kHz bandwidth and NEF with 2.96, which satisfies the requirements for spike (300--7 kHz) recording. Operating from a 1.2V supply, the proposed AFE and DTDM consumes only 2.85 $\mu$W/channel, demonstrating the effectiveness of the chopping and current-reuse techniques.

The DTDM circuit was characterized with different types of neural signals to evaluate signal fidelity. As illustrated in Fig. \ref{fig_Analogresult}(b), the DTDM output exhibits sparse event generation during quiescent periods and adaptive tracking during neural activity. The reconstructed signal from DTDM events achieves a root-mean-square error (RMSE) of 0.0302 compared with the original neural waveform, confirming high-fidelity signal reconstruction. This performance significantly reduces the data bandwidth requirements for wireless transmission in implantable applications.

\begin{figure}[t]
    \centering
    \includegraphics[width=1\linewidth]{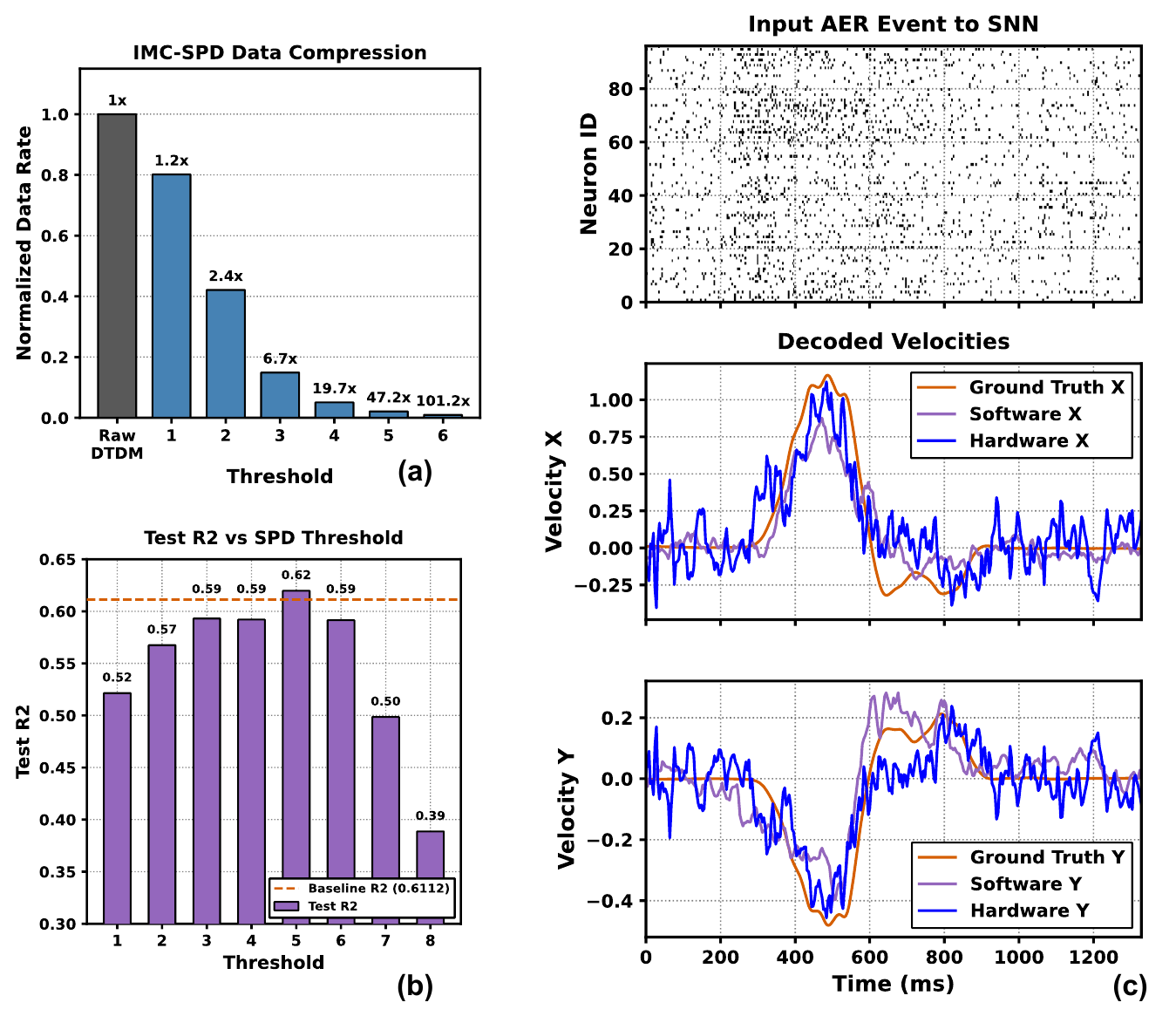}
    \caption{(a) Event rate reduction of IMC SPD. (b) Bi-SNN 2D motor decoding output. (c) Decoding R2 at different spike detection thresholds.}
    \label{fig_Digitalresult}
    \vspace{-10pt}
\end{figure}
\subsection{Spike Detection and Decoding Performance}
We further tested the SoC outputs with in-pixel IMC SPD and the Bi-SNN intention decoder.
Figure \ref{fig_Digitalresult} (a) presents the data rate reduction provided by IMC-SPD at different digital threshold settings. Compared with a raw DTDM frontend, the spike detector reduced the AER event rate by up to 101.2×.
Next in Fig. \ref{fig_Digitalresult} (b) and (c), we tested a 2D velocity decoding task on a primate reaching dataset \cite{o_doherty_2020_3854034,yikNeurobenchFrameworkBenchmarking2025}. We compared system-level decoding accuracy across different threshold settings to demonstrate the trade-off between compression/accuracy in spike detection. The decoding accuracy is comparable with software decoding baselines, reaching $R^2$ of 0.62/0.59 at compression of 47.2/101.2×, respectively.

\begin{table}[t]
\centering
\caption{Comparison of Neural Sensing and Processing SoCs}
\label{soc_comp}
\renewcommand{\arraystretch}{1.3} 
\resizebox{\linewidth}{!}{%
\begin{tabular}{|c|c|c|c|c|c|c|c|}
\hline
& \makecell{TBCAS22 \\ {\cite{anPowerEfficientBrainMachineInterface2022}}} & \makecell{ISSCC23 \\ {\cite{SciCNNTsai}}} & \makecell{ISSCC24 \\ {\cite{shaeri333MiBMI1922024}}} & \makecell{ISSCC24 \\ {\cite{zhong332Sub1mJClass2024}}} & \makecell{ESSERC24 \\ \cite{10719405}} & \makecell{VLSI25 \\ {\cite{alex32Channel196$boldsymbolmuMathbfW$2025}}} & This Work \\ \hline
\textbf{Technology (nm)} & 180 & 40 & 65 & 65 & 180 & 65 & 65 \\ \hline
\textbf{\makecell{Supply Voltage \\ (V)}} & 0.625 & 1.1/0.9 & 1.2 & 1/0.8 & 1.2 & 1.2/0.85 & 1.2/1.1 \\ \hline
\textbf{\makecell{No. of \\ Channels}} & 93 & 22 & 192 & 16 & 1 & 32 & 32 \\ \hline
\textbf{Bandwidth(Hz)} & 300-1k & - & 300-10k & 0.05-400 & 2.5k & 1-500 & 239-7.2k \\ \hline
\textbf{ADC} & - & SAR & SAR & SAR & SAR & SAR & \textbf{DTDM} \\ \hline
\textbf{Pre-Processing} & \makecell{Spike Band \\ Power} & \makecell{Feature \\ Extraction} & Spike Rate & IIR BP & - & \makecell{Feature \\ Extraction} & IMC-SPD \\ \hline
\textbf{\makecell{Decoding \\ Task}} & \makecell{Finger \\ Movement} & \makecell{Seizure \\ Detection} & Handwriting & \makecell{EEG \\ Decoding} & - & \makecell{Psychiatric \\ Classification} & \makecell{Primate \\ Reaching} \\ \hline
\textbf{\makecell{Decoding \\ Model}} & SSKF & SciCNN & DNC+LDA & CNN & - & NAM & Bi-SNN \\ \hline
\textbf{\makecell{Frontend \\ Power/Ch. ($\mu$W)}} & - & 10.62 & 3.44 & 8 & 5.3 & 2.82 & 2.85 \\ \hline
\textbf{\makecell{SoC Power/Ch. \\ ($\mu$W)}} & 135 & 59 & 3.88 & 5.65 & - & 6.14 & \textbf{3.53} \\ \hline
\textbf{\makecell{SoC Area/Ch. \\ (mm$^2$)}} & 0.096 & 0.114 & 0.0105 & 0.46 & - & 0.17 & \textbf{0.034} \\ \hline
\end{tabular}%
}
\end{table}
\section{Conclusion}
In this work, we presented a 32-channel fully event-based SoC for high-density wireless iBMI systems. 
Compared with SOTA neural decoding SoCs (Table \ref{soc_comp}), our test chip achieved ultra-low power consumption of 3.53 $\mu$W per channel and a compact area of 0.034 mm\textsuperscript{2} per channel. 
Our DTDM frontend achieves 26× compression of spike waveform and in-memory spike detection brought about a further 47.2-101.2× event rate reduction. The SoC was validated on a primate reaching task and achieved decoding $R^2$ of 0.62, demonstrating a compact, energy-efficient solution for next-generation implantable BMI systems.

\bibliographystyle{IEEEtran}
\bibliography{ref.bib}

\end{document}